# An Agentic Just-in-Time Adaptive Intervention System for Personalized Sleep Support: Proof-of-Concept Study with N of 1 Data

Nick Rezaee
Department of Biostatistics
Harvard T.H. Chan School of Public Health
Boston, Massachusetts, USA

Chelsea Boccagno
Department of Epidemiology
Harvard T.H. Chan School of Public Health
Boston, Massachusetts, USA

Background: Just-in-time adaptive interventions (JITAIs) can use behavioral data to adapt support to changing contexts, but many rely on predefined rules and manual configuration.

Objective: We developed a proof-of-concept sleep JITAI using an AI agent to review personal data, evaluate reminders, adapt interventions, and record decisions for human review.

Methods: Running in Home Assistant on a configurable schedule, the agent follows a reusable skill file to review 30 days of sleep and behavioral data, including physical activity, smartphone use, and bedtime routines, to identify patterns and create or update automated reminders.

Results: Initial runs demonstrated technical feasibility, successfully completing data review and intervention decisions while limiting reminders to three per day and saving decision records.

Conclusions: Agentic AI may enable flexible, adaptive sleep JITAIs. The architecture supports future comparison with fixed or rule-based interventions, requires human oversight, and could extend to other health behaviors.

## 1 INTRODUCTION

Just-in-time adaptive interventions (JITAIs) are designed to provide support at moments when an individual is most likely to benefit from it [10]. Rather than delivering the same intervention at fixed intervals, JITAIs use information about a person's current state or context to determine whether, when, and how an intervention should be delivered [10]. The approach grew out of earlier work on adaptive interventions and behavioral decision rules and expanded with smartphones, wearable devices, and continuous sensing technologies [10]. Microrandomized trials have been used to test whether and under what conditions prompts delivered through mobile devices support behavior change, with participants repeatedly assigned to receive different intervention options or no intervention [8]. In addition, JITAIs have been developed and evaluated for a range of health behaviors, including physical activity, mental health, substance use, and sleep [3, 10, 12].

JITAIs commonly use decision rules specified in advance. Researchers define which personal data to consider, when decisions are made, which interventions are available, and the conditions for selecting them [10]. A 2024 systematic review of JITAIs targeting physical health outcomes included 45 studies covering 38 different JITAIs, with 39 studies assessing physical activity. Decision rules were mainly based on if-then conditions, such as sending a prompt after 20 minutes of sedentary behavior or when a person had taken fewer than 100 steps in the previous hour [3]. These explicit rules provide transparency because researchers can identify why an intervention was delivered. They also provide control by allowing researchers to specify when reminders should or should not be sent.

However, designing and maintaining these rules can become more demanding when interventions combine several types of data, such as sleep duration, smartphone use, and bedtime routines. Researchers must specify how these data should be considered together, including what to do when they suggest different actions. A person's routines and responses to reminders may also change, so a rule that was previously useful may become less appropriate. For example, a change in work schedule could require different timing for bedtime reminders. Similarly, if a person repeatedly dismisses a reminder, its timing, content, or frequency may need to change. When a JITAI cannot revise its own rules, researchers may need to review the person's data and make these adjustments manually. Repeating this process across many users could require substantial time and effort.

A practical challenge in developing JITAIs is connecting sensor data, decision-making, and intervention delivery so they can work together over time. This requires a platform that can collect new data, make those data available for review, and deliver reminders when specified conditions are met. Home Assistant is an open-source home automation platform that connects devices and services and uses triggers, optional conditions, and actions to run automations [7]. Although designed for tasks such as controlling lights and thermostats, Home Assistant can also bring together data that change over time and carry out scheduled or event-triggered actions. These capabilities make it a potential platform for prototyping adaptive digital health interventions. In this study, Home Assistant connects personal behavioral data with an AI agent that reviews the data and creates or updates smartphone reminders within predefined limits.

Sleep provides a useful setting for testing an AI agent integrated with Home Assistant to review personal data and adapt reminders, because sleep patterns can vary from night to night within the same person [2]. These variations may relate to physical activity, evening routines, and previous sleep. For example, a meta-analysis of 66 studies found that regular exercise was associated with longer sleep duration, less time needed to fall asleep, and better sleep quality [9]. Changes in sleep and related behaviors provide repeated opportunities to review whether reminders remain relevant. Sleep reminders may need to change as a person's routines evolve or as earlier reminders become less useful or unnecessary. An AI agent that regularly reviews these patterns could potentially tailor support over time beyond what a fixed set of rules allows.

We conducted a proof-of-concept study using behavioral data from the first author to assess whether combining multiple personal data sources with an AI agent could support a JITAI for personalized sleep support. In this prototype, interventions were reminders delivered through smartphone notifications. The AI agent reviewed recent behavioral data and existing reminders, then decided whether to create a new reminder, change a reminder's message or delivery conditions, keep it unchanged, or remove it. Home Assistant stored the data and delivered the reminders. The agent followed a reusable Markdown skill file, a plain-text document containing step-by-step instructions. These instructions specified which data to review, how to consider the user's sleep goal, how to assess existing reminders, and which changes were permitted. The same file guided each run and could be edited by the user or researcher to adjust the agent's behavior without rewriting the underlying software. The agent also saved a record of the data reviewed, patterns identified, decisions made, and reasons for those decisions. This study evaluated whether these components could work together automatically, rather than whether the reminders improved sleep.

## 2 METHOD

### 2.1 System Architecture

The proof-of-concept JITAI was implemented within Home Assistant and required a one-time setup, after which it could run automatically each day without manual initiation. Home Assistant served as the persistent infrastructure for storing behavioral data, executing automations, and delivering interventions [6]. The adaptive decision-making component was implemented using Claude Code, which was automatically scheduled to run each day [1]. However, the architecture is not limited to Claude Code; other AI agents based on large language models could potentially be used, provided they can follow the skill instructions and interact with Home Assistant.

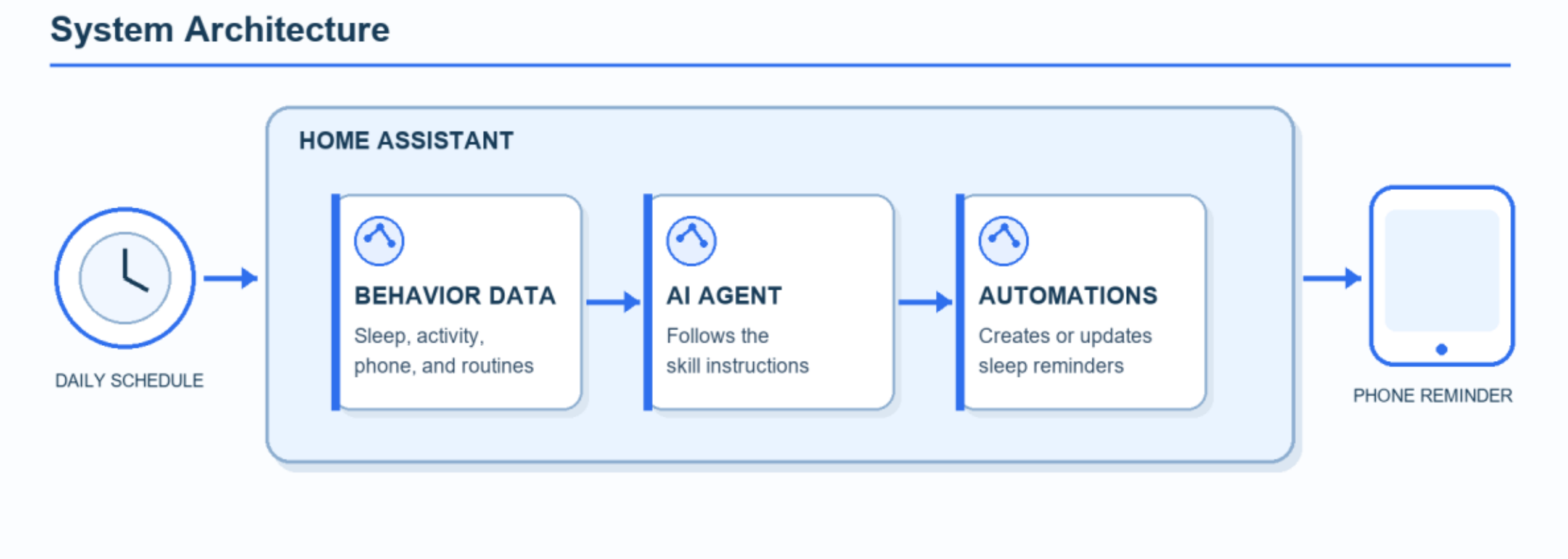

**Figure 1.** This figure illustrates the JITAI architecture for personalized sleep support. At a set time each day, Home Assistant automatically starts the AI agent. Following the skill instructions, the agent reviews sleep data alongside smartphone use and bedtime routines to identify patterns that may be relevant to sleep. The agent then creates or updates reminders when needed, and Home Assistant delivers them as smartphone notifications at the specified times.

### 2.2 Skill-Based Intervention Logic

The agent's behavior was guided by a reusable Markdown skill file, a text document containing step-by-step instructions written in plain language. The file specified which behavioral data to review, how to assess existing smartphone reminders, and when to add, change, keep, or remove a reminder. It also instructed the agent to record each decision and its reasoning so that researchers could review the results of each run.

At each execution, Claude Code was run against this skill file. The same file guided both the analysis of behavioral data and intervention decisions. Because the instructions were stored in a human-readable Markdown file, the decision process could be updated by editing the file rather than changing the underlying system or AI model. The skill defines interventions delivered through smartphone notifications, including what each reminder says, when it is sent, and whether it should be added, changed, kept, or removed.

### 2.3 Behavioral Data and Analysis Window

At each scheduled run, the AI agent reviewed the preceding 30 days of data stored by Home Assistant Recorder, which records sensor readings and events in a database [6]. This window was selected to include recent observations across several weeks. Previous research used 30-day sleep diaries to examine sleep regularity and found that some participants' classifications differed between the first two weeks and the full month [11]. The current implementation includes sleep, smartphone use, television use, and bedtime-routine data. The skill.md file specifies which sensors and data sources the agent should review, such as step counts and smartphone use. Users or researchers can edit the file to change the selected data sources and review a shorter or longer period depending on the target behavior and available data.

The agent reviewed how much data was available and when the data were collected, to identify patterns related to the sleep interventions in the skill. Additionally, the agent considered missing or limited data so that a lack of data did not automatically lead to changes in an intervention.

### 2.4 Intervention Decision Procedure

During the initial setup, the user entered a personal sleep goal to guide the agent's decisions. At each run, the agent reviewed recent behavioral data and existing smartphone reminders in relation to that goal. The agent considered whether enough time and data were available to assess a reminder and whether the observed sleep patterns suggested progress toward the goal. These observations guided whether to add, change, keep, or remove a reminder. Before making changes, the agent checked that no more than three interventions would be delivered per day.

### 2.5 Intervention Delivery

In Home Assistant, automations.yaml stores automation rules in YAML format [4]. Each rule can specify a trigger, optional conditions, and the actions Home Assistant should take [4, 7]. For example, an automation can specify that Home Assistant should send a reminder to the user's phone at a certain time in the evening. In this JITAI, the agent can add, modify, or remove these rules to change how interventions are delivered.

Sleep interventions were set up as Home Assistant automations. In the current prototype, these automations sent bedtime reminders as push notifications to the user's phone through Home Assistant's notification system [5]. Because the interventions run through Home Assistant, they could also use other connected devices and services, such as spoken reminders from smart speakers in the user's home, or lumen changes to smart lights. The ability to use multiple connected devices and services allows the system to deliver support in different ways depending on the user's needs and available devices.

### 2.6 Logging and Auditability

Each time the AI agent ran, it saved a log for later review. The log recorded when the system ran, the period of behavioral data that was examined, how much data was available, and any important patterns or findings identified by the agent. The log also recorded the interventions reviewed, the decision made for each intervention, and the reasoning behind each decision. In addition, the log could include the agent's output, any changes made to the Home Assistant intervention settings, and any errors that occurred during the run. These saved logs allow users and researchers to review the AI agent's past decisions and the reasons behind them. Users or researchers can edit the skill file to specify what the agent records, based on their needs, preferences, or research goals.

### 2.7 Proof-of-Concept Evaluation

The prototype was evaluated by having one of the authors run the system using their own behavioral data. We assessed whether the system could start automatically, retrieve the relevant digital data, and follow the procedures defined in the Markdown skill file. We also assessed whether the AI agent could review existing smartphone reminders, decide whether each reminder should be added, changed, kept, or removed, maintain the limit of three interventions per day, and save a record of each run without manual interaction.

## 3 RESULTS

The system ran eight times over seven consecutive days (August 14–20, 2026) without requiring manual initiation. On each run, it checked the previous 30 days of sleep and behavior data, reviewed the JITAIs already in Home Assistant, determined whether any of the existing JITAIs needed to be changed, and saved a record of what it did. The first author checked the daily runs and their saved records. These tests showed that the system could automatically review data, evaluate reminders, make changes when appropriate, and save a record of each decision. The evaluation focused on whether these steps were completed, rather than whether the reminders improved sleep.

The AI agent made one change during the test period: adding an afternoon walking reminder on August 16. During the first three runs, the agent's saved logs described the available patterns as unclear or supported by too little data to justify a change. On August 16, the agent identified a repeated association between lower daily step counts and shorter sleep duration. Based on this pattern, the agent added the reminder after checking that it did not duplicate an existing reminder. During the next four runs, the agent kept the reminder unchanged, recording that too little time had passed to assess sleep changes and that no clearer pattern had emerged. The agent followed a predefined limit of three interventions per day; zero or one reminder was active during the test period.

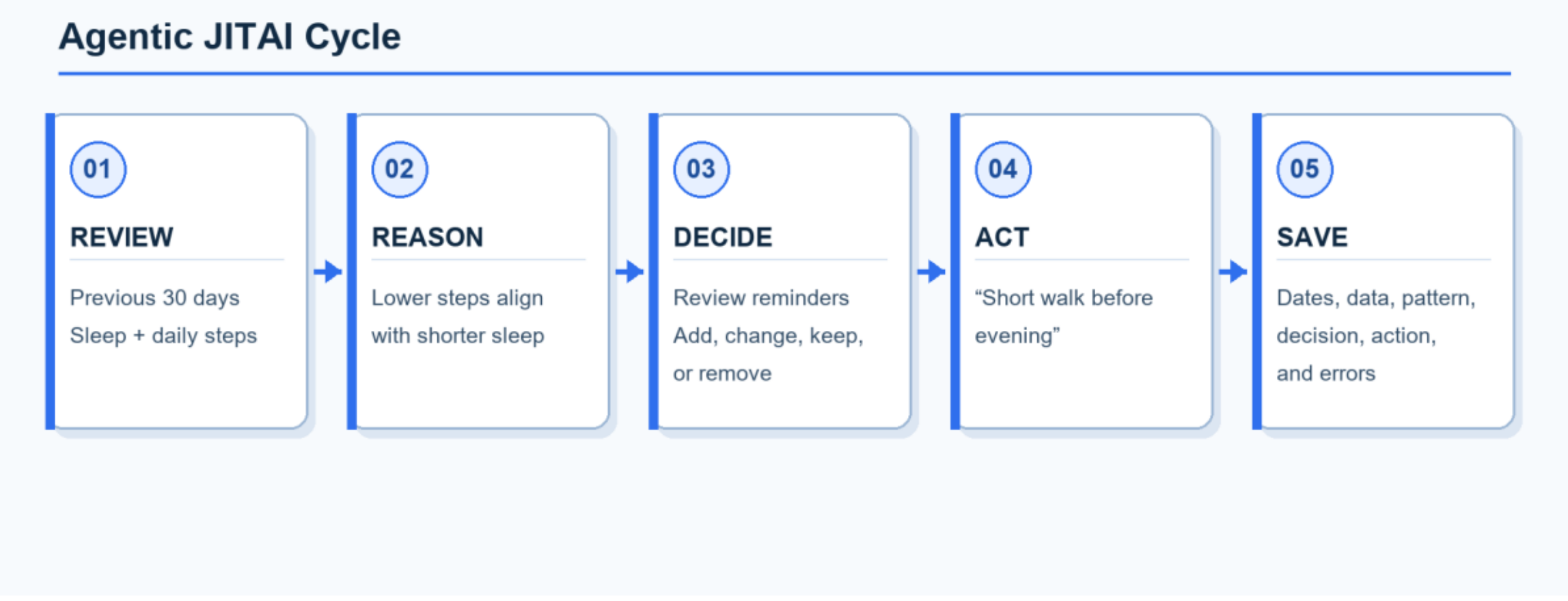

**Figure 2.** This figure illustrates the daily JITAI cycle. Each day, the AI agent (1) reviews recent data, (2) considers behavioral patterns, (3,4) decides whether to add or change a smartphone reminder, and (5) saves a record of the decision.

**Table 1.** Summary of the eight test runs completed automatically by the AI agent. These runs evaluated whether the agent could review data, assess existing smartphone reminders, make changes when appropriate, and record each decision without manual intervention.

| Runs | What the system reviewed | Decision | Action |
|---|---|---|---|
| Aug 14–15 (3) | Last 30 days of sleep and behavior data | No clear pattern | No change |
| Aug 16 (1) | Last 30 days of sleep and behavior data | Repeated association between steps and time asleep | Added an afternoon walking reminder |
| Aug 17–20 (4) | Last 30 days of sleep and behavior data and the walking reminder | No clear pattern to decide to change or add a reminder | No change |

Every run produced a readable saved record, including runs that made no change. Each record listed the dates and amount of data reviewed, missing data, the pattern considered, current reminders, the AI agent's decision, any changes made, and any errors. When the walking reminder shown in Figure 2 was added, the agent checked that the reminder loaded correctly without changing other Home Assistant rules. These records helped distinguish a completed run in which the agent decided no change was needed from a run that failed.

## 4 DISCUSSION

Previous JITAIs commonly use data inputs and decision rules chosen before deployment [3, 10]. In this study, one AI agent completed the full JITAI cycle across eight runs: it reviewed the preceding 30 days of personal data, checked existing reminders, made one change when a repeated pattern appeared, made no change when the evidence was limited, and saved each decision. This work differs from the threshold-based JITAIs described by Fiedler et al. [3] by testing whether the same agent can both (1) interpret updated personal data and (2) directly revise reminders and their delivery conditions in Home Assistant, within human-defined instructions and limits, rather than only applying predefined intervention rules. The main finding is that the system could run as planned; the study does not show that it improved sleep.

Personalized interventions are important because sleep can vary from day to day within the same person [2], and JITAIs are intended to respond to these changes [10]. However, this flexibility also makes the JITAI harder to set up and support across multiple users. Each user needs reliable sensors, clear limits on which data the AI agent can access, secure Home Assistant access, and checks that reminders and saved records work as intended. Supporting more users would require a simpler setup process, consistent ways to organize data from different devices, checks for missing or incorrect data, and separate permissions and records for each person.

The readable skill file allows researchers to adjust the data the agent reviews, the user's sleep goals, and the rules for creating or changing reminders without rewriting the underlying software. This makes the approach easier to adapt to individual needs, but changing the instructions does not guarantee appropriate decisions. Wider use would require testing revised instructions, protecting personal data, and reviewing the agent's decisions. Users should also be able to control data access and pause or stop reminders.

This proof-of-concept study has several limitations. Testing used the first author's personal behavioral data to assess whether the AI agent could complete the full JITAI process before involving additional participants. These real-world data supported an initial technical evaluation within an existing Home Assistant environment. Personal data can help tailor JITAI decisions to an individual's changing circumstances [10] and day-to-day sleep patterns [2]. However, this study focused on whether the data review, reminder updates, and decision logging worked together. Establishing this technical feasibility was a necessary first step before evaluating whether reminders change behavior or improve sleep. Successful execution does not establish that the agent's decisions were appropriate. The agent could act on weak patterns or send reminders that are poorly timed or burdensome. Detailed personal data also raises privacy concerns.

Future studies should test whether agent-selected reminders improve sleep. Comparisons could include reminders selected by an AI agent, reminders adapted through other methods, fixed reminders, and no reminders. Studies involving more people should also assess reliability, user understanding and acceptance, privacy, and user control. Improvements in AI models may support better decisions, but model performance alone cannot establish that an intervention is safe or effective.

The AI agent reviews personal data, checks current smartphone reminders, decides whether a change is needed, updates reminders in Home Assistant, and records the decision. This process could be adapted to support other behaviors, including physical activity, medication routines, stress management, and diet, as a person's needs change over time. Each application would require relevant data, revised skill instructions, suitable safeguards, and testing. Larger and longer studies are needed to determine whether agent-guided JITAIs improve health behaviors or outcomes. The main contribution of this proof of concept is a working approach that brings together diverse personal data in Home Assistant and uses an AI agent to adjust reminder content and timing over time, offering a foundation for more responsive support in personalized medicine.